\documentclass[a4paper,11pt]{article}
\usepackage{pos}

\newcommand{\ctext}[1]{\raise0.2ex\hbox{\textcircled{\scriptsize{#1}}}}

\newcommand{\bbC}{{\mathbb{C}}}
\newcommand{\bbR}{{\mathbb{R}}}

\newcommand{\bc}{\begin{center}}
\newcommand{\ec}{\end{center}}
\newcommand{\br}{\begin{flushright}}
\newcommand{\er}{\end{flushright}}
\newcommand{\bl}{\begin{flushleft}}
\newcommand{\el}{\end{flushleft}}
\newcommand{\ReS}{{\rm Re}\,S}
\newcommand{\ImS}{{\rm Im}\,S}

\title{
Numerical sign problem and the tempered Lefschetz thimble method%
\footnote{Report No.: KUNS-2923}
}
\ShortTitle{
Numerical sign problem and the tempered Lefschetz thimble method
}

\author*[a]{Masafumi Fukuma}
\author[b]{Nobuyuki Matsumoto}
\author[a]{Yusuke Namekawa}

\affiliation[a]{Department of Physics, Kyoto University,\\
  Kitashirakawa-Oiwake-cho, Kyoto 606-8502, Japan}

\affiliation[b]{
  RIKEN/BNL Research center, Brookhaven National Laboratory,\\
  Upton, NY 11973, USA
}

\emailAdd{fukuma@gauge.scphys.kyoto-u.ac.jp}
\emailAdd{nobuyuki.matsumoto@riken.jp}
\emailAdd{namekawa@gauge.scphys.kyoto-u.ac.jp}

\abstract{
The numerical sign problem is a major obstacle 
to the quantitative understanding of many important physical systems 
with first-principles calculations. 
Typical examples for such systems include 
finite-density QCD, 
strongly-correlated electron systems and frustrated spin systems, 
as well as the real-time dynamics of quantum systems. 
In this talk, 
we argue that 
the \emph{tempered Lefschetz thimble method} (TLTM) 
[M.~Fukuma and N.~Umeda, arXiv:1703.00861] 
and its extension, 
the \emph{worldvolume tempered Lefschetz thimble method} (WV-TLTM) 
[M.~Fukuma and N.~Matsumoto, arXiv:2012.08468], 
may be a reliable and versatile solution to the sign problem. 
We demonstrate the effectiveness of the algorithm 
by exemplifying a successful application of WV-TLTM to the Stephanov model, 
which is an important toy model of finite-density QCD. 
We also discuss the computational scaling of WV-TLTM. 
}

\FullConference{%
  Corfu Summer Institute 2021 
  "School and Workshops on Elementary Particle Physics and Gravity"\\
  29 August - 9 October 2021\\
  Corfu, Greece
}

\tableofcontents

\begin{document}
\maketitle

\section{Introduction}
\label{sec:intro}

The numerical sign problem has prevented us 
from the quantitative understanding of many important physical systems 
with first-principles calculations. 
Typical examples for such systems include 
finite-density QCD, 
strongly-correlated electron systems and frustrated spin systems, 
as well as the real-time dynamics of quantum systems. 

The main aim of this talk is to argue 
that the \emph{tempered Lefschetz thimble method} (TLTM) 
\cite{Fukuma:2017fjq} 
and its extension, 
the \emph{worldvolume tempered Lefschetz thimble method} (WV-TLTM) 
\cite{Fukuma:2020fez}, 
may be a reliable and versatile solution to the sign problem. 
The (WV-)TLTM actually has been confirmed to work 
for toy models of some of the systems listed above. 
In this talk, 
we pick up the Stephanov model, 
to which the WV-TLTM is applied. 
This matrix model has played a particularly important role 
in attempts to establish a first-principles calculation method 
for finite-density QCD, 
because it well approximates the qualitative behavior of finite-density QCD 
at large matrix sizes, 
and because it has a serious sign problem 
which had not been solved by other methods than the (WV-)TLTM. 
We also discuss the computational scaling of WV-TLTM.

\section{Sign problem}
\label{sec:sign_problem}

\subsection{What is the sign problem?}
\label{sec:what_is_sign_problem}

Our aim is to numerically estimate the expectation value 
defined in a path-integral form:
\begin{align}
  \langle \mathcal{O}(x) \rangle \equiv
  \frac{\int dx\,e^{-S(x)}\,\mathcal{O}(x)}{\int dx\,e^{-S(x)}}.
\label{vev}
\end{align}
Here, $x=(x^i)\in\bbR$ is a dynamical variable 
of $N$ degrees of freedom (DOF), 
$S(x)$ the action, 
and $\mathcal{O}(x)$ a physical observable of interest. 

When $S(x)$ is real-valued, 
one can regard $p_{\rm eq}(x)\equiv e^{-S(x)}/\int dx\,e^{-S(x)}$ 
as a probability distribution, 
and can estimate $\langle \mathcal{O}(x) \rangle$ 
with a sample average as 
\begin{align}
  \langle\mathcal{O}(x)\rangle \approx 
  \frac{1}{N_{\rm conf}}\,\sum_{k=1}^{N_{\rm conf}} \mathcal{O}(x^{(k)}).
\end{align}
Here, $\{x^{(k)}\}$ is 
a sample (a set of configurations) of size $N_{\rm conf}$, 
that is generated as a suitable Markov chain 
with the equilibrium distribution $p_{\rm eq}(x)$.

The above prescription is no longer applicable
when the action has an imaginary part as 
$S(x)=S_R(x)+i S_I(x)\in\bbC^N$. 
A naive way to handle this 
is the so-called reweighting method, 
where we treat $e^{-S_R(x)}/\int dx\,e^{-S_R(x)}$ as a new weight 
and rewrite the expression \eqref{vev} 
as a ratio of reweighted averages: 
\begin{align}
  \langle \mathcal{O}(x) \rangle =
  \frac{\langle e^{-i S_I(x)}\,\mathcal{O}(x) \rangle_{\rm rewt}}
  {\langle e^{-i S_I(x)}\, \rangle_{\rm rewt}}
  \quad
  \Bigl(
    \langle f(x) \rangle_{\rm rewt}
    \equiv \frac{\int dx\,e^{-S_R(x)}\,f(x)}{\int dx\,e^{-S_R(x)}}
  \Bigr). 
\label{vev-rewt}
\end{align}
However, when the DOF, $N$, is very large, 
the reweighted averages can become 
vanishingly small of $e^{-O(N)}$, 
even though the operator itself is $O(1)$. 
This should not be a problem 
\emph{if} we can estimate 
both the numerator and the denominator precisely. 
However, in the numerical computation, 
they are estimated separately with statistical errors: 
\begin{align}
  \langle \mathcal{O}(x) \rangle \equiv 
  \frac{\langle e^{-i S_I(x)} \mathcal{O}(x) \rangle_{\rm rewt}}
  {\langle e^{-i S_I(x)} \rangle_{\rm rewt}}
  \approx
  \frac{e^{-O(N)} \pm O(1/\sqrt{N_{\rm conf}})}
  {e^{-O(N)} \pm O(1/\sqrt{N_{\rm conf}})}. 
\end{align}
Thus, in order for the statistical errors 
to be smaller than the mean values, 
the sample size must be exponentially large 
with respect to DOF, 
namely, $N_{\rm conf} \gtrsim e^{O(N)}$. 
The need of this unrealistically large numerical cost 
is called the sign problem. 

\subsection{Various approaches proposed so far}
\label{sec:approaches}

We list some of the approaches proposed so far, 
which are intended to solve the sign problem. \vspace{4pt}

\noindent
\underline{\textbf{class 1}: 
no use of reweighting}

A typical algorithm in this class 
is the complex Langevin method 
\cite{Parisi:1984cs,Klauder:1983nn,Klauder:1983sp,
Aarts:2011ax,Aarts:2013uxa,Nagata:2016vkn}, 
where the complex Boltzmann weight 
is rewritten to a positive probability distribution 
over a complex space $\bbC^N$. 
Although its numerical cost is low $[\sim O(N)]$,  
it often exhibits a wrong convergence  
(gives incorrect estimates with small statistical errors) 
at parameter values of physical importance. \vspace{4pt}

\noindent
\underline{\textbf{class 2}: 
deforming the integration surface}

A typical algorithm is the Lefschetz thimble method 
\cite{Witten:2010cx,Cristoforetti:2012su,Fujii:2013sra,
Alexandru:2015sua,Fukuma:2017fjq,Alexandru:2017oyw,
Fukuma:2019wbv,Fukuma:2019uot,Fukuma:2020fez,Fukuma:2021aoo}, 
where the integration surface $\Sigma_0=\bbR^N$ 
is continuously deformed to a new surface $\Sigma_t \subset \bbC^N$.%
\footnote{
  This algorithm will be explained in detail in the next section. 
  Another interesting algorithm is the path-optimization method 
  \cite{Mori:2017pne,Alexandru:2018fqp}, 
  where the integration surface is looked for 
  with the machine learning technique 
  so that the average phase factor is maximized. 
} 
The flow time $t$ is taken sufficiently large 
so that $\Sigma_t$ is close to a union of Lefschetz thimble, 
$\bigcup_\sigma \mathcal{J}_\sigma$, 
on each of which $\ImS(z)$ $(z\in\mathcal{J}_\sigma)$ is constant. 

Generic Lefschetz thimble method has been shown 
to suffer from an ergodicity problem 
for physically important parameter regions 
of a model \cite{Fujii:2015bua}, 
where multiple thimbles become relevant 
that are separated by infinitely high potential barriers.  
This problem was resolved by tempering the system with the flow time 
\cite{Fukuma:2017fjq}.%
\footnote{
  A similar idea is proposed in Ref.~\cite{Alexandru:2017oyw}.
} 
This \emph{tempered Lefschetz thimble method} (TLTM) 
solves 
both the sign problem (serious at small flow times) 
and the ergodicity problem (serious at large flow times) 
simultaneously. 
The disadvantage is its high numerical cost of $O(N^{3-4})$. 
Recently, this numerical cost has been substantially reduced 
[expected to be $O(N^{\sim 2.25})$] 
with a new method, 
the \emph{worldvolume tempered Lefschetz thimble method} (WV-TLTM), 
which is based on the idea 
to perform the Hybrid Monte Carlo 
on a continuous accumulation of deformed integration surfaces 
(the worldvolume) \cite{Fukuma:2020fez}. 
The algorithm (WV-)TLTM is the main subject in this talk.%
\footnote{
  See Ref.~\cite{Alexandru:2020wrj} for a review from a different viewpoint.
} 
\vspace{4pt}

\noindent
\underline{\textbf{class 3}: 
no use of MC in the first place}

A typical algorithm in this class is the tensor network method 
(especially the tensor renormalization group method 
\cite{Levin:2006jai}).%
\footnote{
  See Ref.~\cite{Fukuma:2021cni} for a recent attempt 
  to apply the tensor renormalization group method 
  to Yang-Mills theory.
} 
This is good at calculating the free energy 
in the thermodynamic limit, 
but not so much efficient to calculate correlation functions 
at large distances. 
We expect this method to play a complementary role 
to methods based on Markov chain Monte Carlo.

\section{Lefschetz thimble method}
\label{sec:Lefschetz_thimble}


We complexify the dynamical variable $x=(x^i)\in\bbR^N$ 
to $z=(z^i=x^i+i y^i)\in\bbC^N$. 
We set an assumption (which holds for most cases) 
that 
$e^{-S(z)}$ and $e^{-S(z)} \mathcal{O}(z)$ are 
entire functions over $\bbC^N$. 
Then, Cauchy's theorem ensures that 
the integrals do not change their values 
under continuous deformation of the integration surface: 
$\Sigma_0=\bbR^N \to \Sigma\,(\subset \bbC^N)$, 
where the boundary at $|x|\to\infty$ is fixed 
so that the convergence of integration holds under the deformation: 
\begin{align}
  \langle \mathcal{O}(x) \rangle 
  = \frac{\int_{\Sigma_0} dx\,e^{-S(x)}\,\mathcal{O}(x)}
  {\int_{\Sigma_0} dx\,e^{-S(x)}}
  = \frac{\int_\Sigma dz\,e^{-S(z)}\,\mathcal{O}(z)}
  {\int_\Sigma dz\,e^{-S(z)}}.
\label{vev2}
\end{align}
Thus, even when the sign problem is severe 
on the original surface $\Sigma_0$, 
it will be significantly reduced 
if $\ImS(z)$ is almost constant on the new surface $\Sigma$.

The prescription for the deformation is given 
by the anti-holomorphic gradient flow: 
\begin{align}
  \dot{z}_t = \overline{\partial S(z_t)}
  \quad \mbox{with} \quad
  z_{t=0}=x.
\label{flow}
\end{align}
The most important property of this flow equation is 
the following inequality: 
\begin{align}
  [S(z_t)]^{\cdot} = \partial S(z_t)\cdot \dot{z}_t
  = |\partial S(z_t)|^2 \geq 0,
\end{align}
from which we find that 

\noindent
(i)~~%
$\ReS(z_t)$ always increases along the flow 
except at critical points,%
\footnote{
  $\zeta$ is said to be a critical point 
  when $\partial S(\zeta)=(\partial_i S(\zeta)) =0$.
} 

\noindent
(ii)~%
$\ImS(z_t)$ is constant along the flow.

The Lefschetz thimble $\mathcal{J}$ 
associated with a critical point $\zeta$ 
is defined by a set of orbits starting at $\zeta$. 
From this construction and property (ii), 
we easily see that $\ImS(z)$ is constant on $\mathcal{J}$ 
[i.e., $\ImS(z)=\ImS(\zeta)~(z\in\mathcal{J})$]. 
Denoting the solution of Eq.~\eqref{flow} by $z_t(x)$ 
and assuming that 
$\Sigma_t\equiv \{z_t(x)|\, x\in\bbR^N\}$ 
approaches a single Lefschetz thimble $\mathcal{J}$, 
we expect that the sign problem disappears on $\Sigma_t$ 
if we choose a sufficiently large $t$. 

Let us see
how the sign problem disappears as flow time $t$ increases. 
The integrals on a deformed surface $\Sigma_t$  
can be rewritten as 
\begin{align}
  \langle \mathcal{O}(x) \rangle 
  = \frac{\langle e^{i \phi(z)} \mathcal{O}(z) \rangle_{\Sigma_t}}
  {\langle e^{i \phi(z)} \rangle_{\Sigma_t}},
\label{vev3}
\end{align}
where%
\footnote{
  Note that 
  $\langle f(z) \rangle_{\Sigma_0} = \langle f(x) \rangle_{\rm rewt}.$
} 
\begin{align}
  \langle f(z) \rangle_{\Sigma_t} 
  \equiv \frac{\int_{\Sigma_t} |dz|\,e^{-\ReS(z)} f(z) }
  {\int_{\Sigma_t} |dz|\,e^{-\ReS(z)}},
  \quad
  e^{i \phi(z)} \equiv e^{-i \ImS(z)}\,\frac{dz}{|dz|}.
\end{align}
As can be easily checked for a Gaussian case, 
the integrals take the form $O(e^{-e^{-\lambda t}O(N)})$, 
where $\lambda$ is a typical singular value 
of $\partial_i \partial_j S(\zeta)$. 
Thus, the numerical estimate now becomes 
\begin{align}
  \langle \mathcal{O}(x) \rangle 
  \approx
  \frac{O(e^{-e^{-\lambda t}O(N)}) \pm O(1/\sqrt{N_{\rm conf}})}
  {O(e^{-e^{-\lambda t}O(N)}) \pm O(1/\sqrt{N_{\rm conf}})}, 
\end{align}
from which we see that the main parts become $O(1)$ 
when the flow time $t$ satisfies 
a relation $e^{-\lambda t}O(N)=O(1)$. 
We thus see that the sign problem disappears 
at flow times $t\gtrsim T=O(\ln N)$.

\section{Tempered Lefschetz thimble method (TLTM)}
\label{sec:TLTM}

\subsection{Ergodicity problem in the original Lefschetz thimble method}
\label{sec:ergodicity_problem}

So far, so good; 
when a single Lefschetz thimble is relevant to estimation,
one can resolve the sign problem 
simply by taking a sufficiently large flow time. 
However, this nice story no longer holds true 
when multiple thimbles are involved in estimation, 
because there comes up another problem (ergodicity problem) 
as the flow time increases.

\begin{figure}[ht]
  \centering
  \includegraphics[width=70mm]{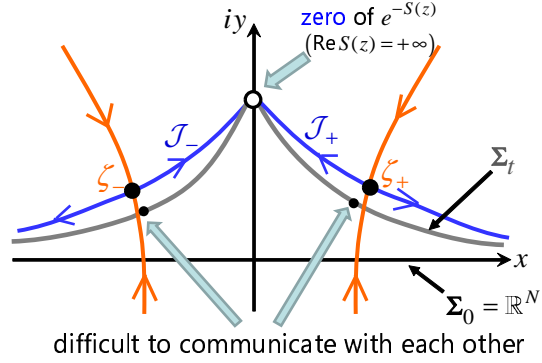}
  \caption{Ergodicity problem.}
  \label{fig:ergodicity_problem}
\end{figure}

Figure \ref{fig:ergodicity_problem} describes the case 
$e^{-S(x)}=e^{-\beta x^2/2}\,(x-i)^\beta$ $(\beta\gg 1)$. 
In addition to two critical points 
$\zeta_\pm=\pm \sqrt{3}/2 +(1/2)\,i$ 
and the associated Lefschetz thimbles $\mathcal{J}_\pm$, 
here is the zero of $e^{-S(z)}$ at $z=i$. 
We see that the integration surface $\Sigma_T$ 
is separated into two parts 
by an infinitely high potential barrier at the zero. 
It is thus very hard 
for two configurations on different parts to communicate 
in stochastic processes, 
which means that it takes a very long computation time 
for the system to reach equilibrium. 
%

\subsection{Basic algorithm of TLTM}
\label{sec:basic_algorithm_TLTM}

The tempered Lefschetz thimble method \cite{Fukuma:2017fjq}
was invented to overcome this problem 
by implementing the tempering algorithm 
\cite{Marinari:1992qd,Swendsen1986,Geyer1991,Hukushima1996}
to the thimble method, 
where the flow time is used as a tempering parameter 
(see Fig.~\ref{fig:tltm}). 
The basic algorithm goes as follows:\vspace{4pt}
\begin{figure}[ht]
  \centering
  \includegraphics[width=70mm]{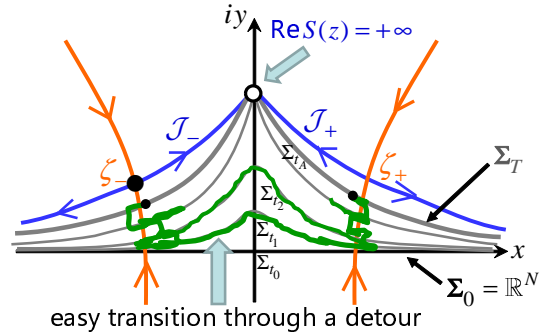}
  \caption{Tempered Lefschetz thimble method (TLTM).}
  \label{fig:tltm}
\end{figure}%

\noindent \underline{Step 0.}\hspace{2mm}%
We fix the target flow time $T$ 
so that the sign problem is not serious 
for a sample on $\Sigma_T$ 
except for the ergodicity problem. 
This is judged by looking at the average phase factor 
$|\langle e^{i\phi(z)} \rangle_{\Sigma_T}|$.\vspace{2pt}

\noindent \underline{Step 1.}\hspace{2mm}%
We introduce replicas in between 
the initial integration surface $\Sigma_0=\bbR^N$ 
and the target deformed surface $\Sigma_T$ as 
$\{\Sigma_{t_0=0},\Sigma_{t_1},\ldots,\Sigma_{t_A=T}\}$.\vspace{2pt}

\noindent \underline{Step 2.}\hspace{2mm}%
We set up a Markov chain for the extended configuration space 
$\{(x,t_\alpha) |\,x\in\bbR^N,\,\alpha=0,1,\ldots,A\}$.\vspace{2pt}

\noindent \underline{Step 3.}\hspace{2mm}%
After equilibration, 
we estimate observables with a sample on $\Sigma_T$.
\vspace{4pt}

This tempering method prompts the equilibration on $\Sigma_T$ 
because two configurations on different connected components 
now can communicate easily by passing through a detour. 
Thus, the TLTM solves 
both the sign and ergodicity problems simultaneously. 

\subsection{Comment on transitions between adjacent replicas}
\label{sec:comment_on_transitions}

We here comment that 
one can expect a significant acceptance rate 
for transitions between adjacent replicas \cite{Fukuma:2019wbv}. 
To see this, let us use the initial configurations $x\in\bbR^N$ 
as common coordinates for different replicas. 
When we employ the simulated tempering \cite{Marinari:1992qd} 
for a tempering method as in the previous subsection, 
a configuration $(x,t_\alpha)$ 
moves to $(x,t_{\alpha\pm 1})$ 
(after it explores on $\Sigma_{t_\alpha}$), 
keeping the $x$-coordinate values the same.%
\footnote{
  When the parallel tempering 
  \cite{Swendsen1986,Geyer1991,Hukushima1996} 
  is employed as in Ref.~\cite{Fukuma:2017fjq}, 
  two configurations on adjacent replicas, 
  $(x,t_\alpha)$ and $(x',t_{\alpha+1})$, 
  move as 
  $(x,t_\alpha)\to (x,t_{\alpha+1})$ 
  and $(x',t_{\alpha+1})\to (x',t_\alpha)$, 
  again keeping the $x$-coordinate values the same.
} 
Since the probability distribution on every replica
has peaks at the same points $x_\sigma$, 
where $x_\sigma$ flows to a critical point $z_\sigma$, 
we can expect a significant overlap 
between distributions on two adjacent replicas. 

\subsection{Computational cost for the original TLTM}
\label{sec:computational_cost_TLTM}

An obvious advantage of the original TLTM is its versatility; 
the method can be applied to any system 
once it is formulated in a path-integral form
with continuous variables, 
resolving the sign and ergodicity problems simultaneously. 
A disadvantage is its high numerical cost. 
It is expected to be $O(N^{3-4})$ 
due to (a) the increase of the necessary number of replicas 
[probably as $O(N^{0-1})$] 
and (b) the need to compute the Jacobian matrix of the flow, 
$J(x)\equiv (\partial z^i_t(x)/\partial x^a)$, 
every time we move configurations between adjacent replicas 
[$O(N^3)$]. 
The worldvolume TLTM \cite{Fukuma:2020fez} was introduced 
to significantly reduce the computational cost.

\section{Worldvolume tempered Lefschetz thimble method (WV-TLTM)}
\label{sec:WV-TLTM}

\subsection{Basic idea of the Worldvolume TLTM}
\label{sec:basic_idea_WV-TLTM}

\begin{figure}[ht]
  \centering
  \includegraphics[width=70mm]{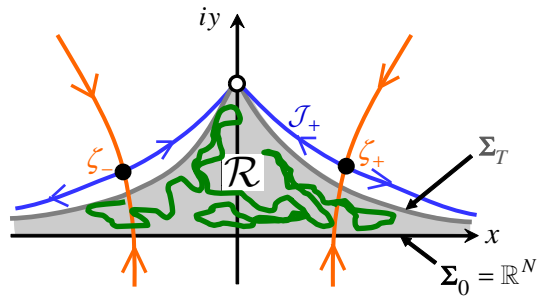}
  \caption{Worldvolume $\mathcal{R}$ of WV-TLTM.}
  \label{fig:wv-tltm}
\end{figure}

Instead of introducing a finite set of replicas 
(a finite set of integrations surfaces), 
we consider in the WV-TLTM 
a HMC algorithm 
on a continuous accumulation of deformed integration surfaces, 
\begin{align}
  \mathcal{R}\equiv \bigcup_{0\leq t\leq T}\Sigma_t 
  = \bigl\{ z_t(x) \big|\,t \in [0,T],~x \in \bbR^N \bigr\}. 
\end{align}
We call $\mathcal{R}$ the \emph{worldvolume} 
because this is an orbit of integration surface 
in the ``target space'' $\bbC^N=\bbR^{2N}$ 
(see Fig.~\ref{fig:wv-tltm}).%
\footnote{
  We here use a terminology in string theory, 
  where an orbit of particle is called a worldline,
  that of string a worldsheet, 
  and that of membrane (surface) a worldvolume.
} 

Keeping the original virtues intact 
(solving the sign and ergodicity problems simultaneously), 
the new algorithm significantly reduces the computational cost. 
In fact, we no longer need to introduce replicas explicitly 
or to calculate the Jacobian matrix 
in every molecular dynamics process, 
and we can move configurations largely 
due to the use of HMC algorithm. 

The key idea behind the algorithm is again Cauchy's theorem. 
We start from the expression \eqref{vev2}:
\begin{align}
  \langle \mathcal{O}(x) \rangle 
  = \frac{\int_{\Sigma_0} dx\,e^{-S(x)}\,\mathcal{O}(x)}
  {\int_{\Sigma_0} dx\,e^{-S(x)}}
  = \frac{\int_{\Sigma_t} dz_t\,e^{-S(z_t)}\,\mathcal{O}(z_t)}
  {\int_{\Sigma_t} dz_t\,e^{-S(z_t)}}.
\label{vev2a}
\end{align}
Cauchy's theorem ensures that both the numerator and the denominator 
do not depend on $t$, 
so that we can average over $t$ with an arbitrary weight $e^{-W(t)}$, 
leading to an integration over $\mathcal{R}$:%
\footnote{
  The weight $e^{-W(t)}$ is determined 
  such that the probability to appear on $\Sigma_t$ 
  is (almost) independent of $t$. 
} 
\begin{align}
  \langle \mathcal{O}(x) \rangle 
  = \frac{\int_0^T dt\,e^{-W(t)}
  \int_{\Sigma_t} dz_t\,e^{-S(z_t)} \mathcal{O}(z_t)}
  {\int_0^T dt\,e^{-W(t)}\int_{\Sigma_t} dz_t\,e^{-S(z_t)}}
  = \frac{\int_\mathcal{R} dt\,dz_t\,
  e^{-W(t)-S(z_t)} \mathcal{O}(z_t)}
  {\int_\mathcal{R} dt\,dz_t\,e^{-W(t)-S(z_t)}}.
\end{align}

\subsection{Algorithm}
\label{sec:WV-TLTM_algorithm}

An explicit implementation can go in two ways, 
as described in the original paper \cite{Fukuma:2020fez}. 
One is the \emph{target-space picture}, 
in which the HMC is performed on the worldvolume $\mathcal{R}$ 
that is treated as a submanifold in the target space $\bbC^N$. 
The other is the \emph{parameter-space picture}, 
in which the HMC is performed on the parameter space $\{(x,t)\}$.%
\footnote{
  The latter picture was further studied in Ref.~\cite{Fujisawa:2021hxh}.
  In this picture, however, 
  the Jacobian determinant $\det J(x)$ 
  is treated as part of observable, 
  which is exponentially large 
  and has no guarantee to have a significant overlap 
  with the weight $e^{-\ReS(z_t(x))}$. 
  This is why we have not pursued the second option seriously 
  in the original paper \cite{Fukuma:2020fez}. 
} 

In the target-space picture, 
we first parametrize the induced metric on $\mathcal{R}$ 
with the ADM decomposition \cite{Arnowitt:1962hi}: 
\begin{align}
  ds^2 = \alpha^2\,dt^2
  + \gamma_{ab}\,(dx^a + \beta^a\,dt)\,(dx^b + \beta^b\,dt).
\end{align}
Here, the functions $\alpha$ and $\beta^a$  
are called the lapse and the shifts, respectively, 
and $\gamma_{ab}$ is the induced metric on $\Sigma_t$. 
The invariant volume element on $\mathcal{R}$ is then given by 
\begin{align}
  Dz = \alpha\,dt\,|dz_t(x)| = \alpha \,|\det J|\,dt\,dx 
  \quad
  \bigl(|\det J|=\sqrt{\det \gamma}\bigr),
\end{align}
and the expectation value can be rewritten 
to a ratio of reweighted averages on $\mathcal{R}$: 
\begin{align}
  \langle \mathcal{O}(x) \rangle = 
  \frac{\int_\mathcal{R} Dz\,e^{-V(z)}\,A(z)\,\mathcal{O}(z)}
  {\int_\mathcal{R} Dz\,e^{-V(z)}\,A(z)}
  = \frac{\langle A(z)\,\mathcal{O}(z) \rangle_\mathcal{R}}
  {\langle A(z) \rangle_\mathcal{R}}.
\end{align}
Here, the reweighted average of a function $f(z)$ is defined by 
\begin{align}
  \langle f(z) \rangle_\mathcal{R}
  \equiv \frac{\int_\mathcal{R} Dz\,e^{-V(z)}\,f(z)}
  {\int_\mathcal{R} Dz\,e^{-V(z)}}
\end{align}
with $V(z) \equiv \ReS(z)+W(t(z))$, 
and the associated reweighting factor takes the form 
\begin{align}
  A(z) \equiv \frac{dt\,dz_t}{Dz}\,e^{-i\ImS(z)}
  = \alpha^{-1}(z)\,\frac{\det J}{|\det J|}\,e^{-i\ImS(z)}.
\end{align}

\begin{figure}[t]
  \centering
  \includegraphics[width=70mm]{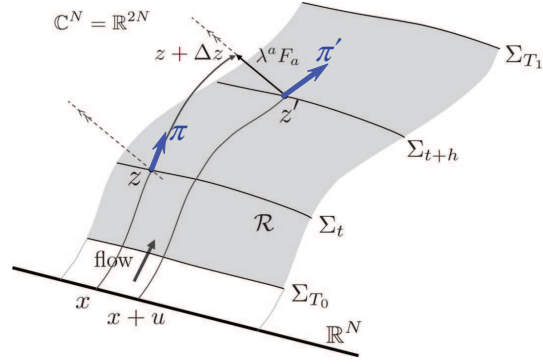}
  \caption{RATTLE on the worldvolume $\mathcal{R}$ \cite{Fukuma:2020fez}.}
  \label{fig:rattle}
\end{figure}

The reweighted average can be estimated with the RATTLE algorithm 
\cite{Andersen:1983,Leimkuhler:1994}, 
where molecular dynamics is performed on $\mathcal{R}$ 
which is treated as a submanifold of $\bbC^N$ \cite{Fukuma:2020fez}. 
The algorithm takes the following form 
(see Fig.~\ref{fig:rattle}):%
\footnote{
  RATTLE on a single Lefschetz thimble $\mathcal{J}=\Sigma_{t=\infty}$ 
  was first introduced in Ref.~\cite{Fujii:2013sra}, 
  which is extended to $\Sigma_t$ with finite $t$ 
  in Ref.~\cite{Alexandru:2019} 
  (see also Ref.~\cite{Fukuma:2019uot} for the combination of RATTLE 
  with the tempering algorithm).  
} 
\begin{align}
  \pi_{1/2} &= \pi - \Delta s\,\bar\partial V(z) - \lambda^a F_a(z),
\\
  z' &= z + \Delta s\,\pi_{1/2},
\\
  \pi' &= z -\Delta s\,\bar\partial V(z') - \lambda'^{a} F_a(z').
\end{align}
Here, $F_a(z) \equiv i J_a(z)$ $(a=1,\ldots,N)$
with $J_a \equiv (J^i_a=\partial z^i_t(x)/\partial x^a)$ 
form a basis of the normal space $N_z\Sigma_t$ 
at $z\in\Sigma_t\,(\subset\mathcal{R})$. 
The Lagrange multipliers $\lambda^a$ and $\lambda'^{a}$ 
are determined using $E_0(z)\equiv\overline{\partial S(z)}$ 
such that 
\begin{align}
  \bullet&~~
  z'\in\mathcal{R}~~~\mbox{and}~~~
  \lambda^a F_a(z) \perp E_0(z),
\\
  \bullet&~~
  \pi' \in T_{z'}\mathcal{R}~~~\mbox{and}~~~
  \lambda'^a F_a(z') \perp E_0(z').
\end{align}
The second equation in each line 
ensures that $\lambda^a F_a(z)$ actually 
belongs to $N_z\mathcal{R}\,(\subset N_z \Sigma_t)$. 
The statistical analysis method for WV-TLTM 
(or more generally, for WV-HMC 
that is the HMC algorithm on a foliated manifold) 
is established in Ref.~\cite{Fukuma:2021aoo}. 

\subsection{Various models to which (WV-)TLTM is applied}
\label{sec:various_models}

The (WV-)TLTM has been successfully applied to various models, 
including \\
\noindent
$\bullet$~~%
$(0+1)$-dimensional massive Thirring model \cite{Fukuma:2017fjq}\\
\noindent
$\bullet$~~%
two-dimensional Hubbard model \cite{Fukuma:2019wbv,Fukuma:2019uot}\\
\noindent
$\bullet$~~%
Stephanov model 
(a chiral random matrix model as a toy model of finite density QCD)
\cite{Fukuma:2020fez}\\
\noindent
$\bullet$~~%
antiferromagnetic Ising model on the triangular lattice 
\cite{Fukuma:2020JPS}. \\
Correct results have always been obtained 
when they can be compared with analytic results, 
although the system sizes are yet small. 

In the next section, 
we discuss the application of WV-TLTM to the Stephanov model.

\section{Application to the Stephanov model}
\label{sec:application_Stephanov_model}

\subsection{Stephanov model}
\label{sec:Stephanov_model}

The finite density QCD is given 
by the following partition function 
after $N_f$ quark fields 
(assumed to have the same mass) are integrated out: 
\begin{align}
  Z_{\rm QCD} &= {\rm tr}\,e^{-\beta(H-\mu N)}
\nonumber\\
  &= \int [dA_\mu]\,e^{(1/2g_0^2)\,\int d^4 x\,{\rm tr}\,F_{\mu\nu}^2}\,
  {\rm Det}\,{}^{N_f}
  \left(\begin{array}{cc}
    m & \sigma_\mu (\partial_\mu + A_\mu) + \mu \\
    \sigma^\dag (\partial_\mu+A_\mu) + \mu & m
  \end{array}\right).
\end{align}
The Stephanov model \cite{Stephanov:1996ki,Halasz:1998qr} 
takes the following form at temperature $T=0$: 
\begin{align}
  Z_{\rm Steph} = \int d^2 W\,e^{-n\,{\rm tr}\,W^\dag W}\,
  \det{}^{N_f} 
  \left(\begin{array}{cc}
    m & i W + \mu \\
    i W^\dag + \mu & m
  \end{array}\right),
\end{align}
where the $n\times n$ complex matrix $W=(W_{ij})=(X_{ij}+i Y_{ij})$ 
represents quantum-field degrees of freedom 
(including space-time dependences).%
\footnote{
  The degrees of freedom is given by $N=2n^2$, 
  which should be compared with those of link variables, 
  $4L^4 (N_c^2-1)$, 
  where $L$ is the linear size of four-dimensional square lattice 
  and $N_c$ is color.
} 
This model plays a particularly important role 
because 
(a) it well approximates the qualitative behavior of finite-density QCD 
at large matrix sizes 
and 
(b) it has a serious sign problem 
which can hardly be solved by the complex Langevin method 
due to a wrong convergence \cite{Bloch:2017sex}. 

\begin{figure}[ht]
  \centering
  \includegraphics[width=140mm]{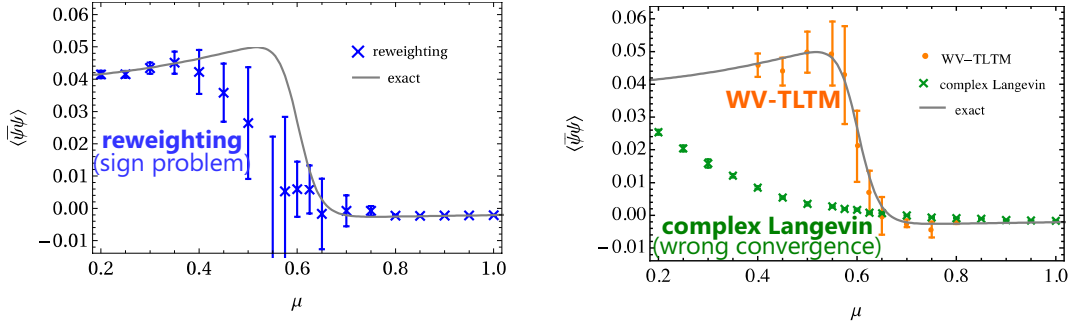}
  \caption{Chiral condensate $\langle\bar\psi\psi\rangle
  \equiv (1/2n)(\partial/\partial m)\ln Z_{\rm Steph}$ 
  \cite{Fukuma:2020fez}.}
  \label{fig:chiral_condensate}
\end{figure}

\begin{figure}[ht]
  \centering
  \includegraphics[width=140mm]{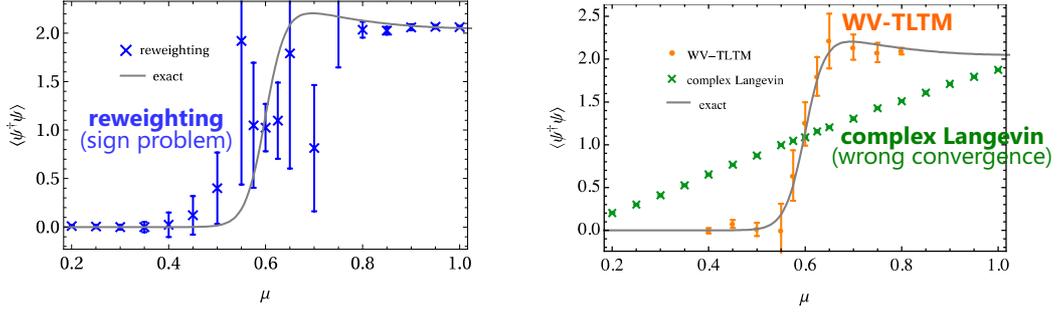}
  \caption{Number density $\langle\psi^\dag\psi\rangle
  \equiv (1/2n)(\partial/\partial \mu)\ln Z_{\rm Steph}$
  \cite{Fukuma:2020fez}.}
  \label{fig:number_density}
\end{figure}

Figures \ref{fig:chiral_condensate} 
and \ref{fig:number_density} show the results 
for the chiral condensate $\langle \bar\psi \psi \rangle$ 
and the number density $\langle \psi^\dag \psi \rangle$ 
at $n=10$, $m=0.004$ and $N_f=1$ 
obtained with the WV-TLTM, 
where the sample size is $N_{\rm conf} = 4,000 - 17,000$ 
(varying on $\mu$). 
We see that they agree with the exact results within statistical errors. 
For comparison, 
we also plot the results obtained 
with the naive reweighting method 
(showing large deviations from the exact values due to the sign problem)
and also with the complex Langevin method 
(exhibiting a serious wrong convergence). 
The sample size is $N_{\rm conf} = 10^4$ 
for both the reweighting and the complex Langevin.
%

\subsection{Computational scaling}
\label{sec:computational_scaling}

In the RATTLE algorithm, 
we need to make an inversion of the linear problem, 
$J v = b$ 
($J$: the Jacobian matrix). 
The total numerical cost of WV-TLTM depends on 
which solver is used. 

\begin{figure}[ht]
  \centering
  \includegraphics[width=90mm]{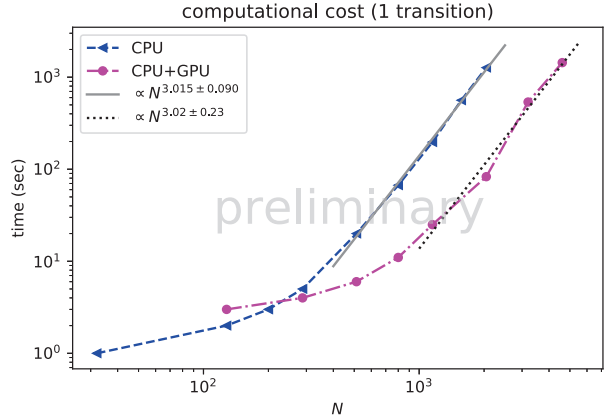}
  \caption{Computation time to generate a configuration 
  with a direct method in the linear inversion.}
  \label{fig:comp_cost}
\end{figure}

When a direct method (e.g., LU decomposition) is used, 
the computational cost is expected to be $O(N^3)$.  
In this case, 
the Jacobian matrix $J=(J_t(x))$ is explicitly computed  
by numerically integrating the differential equation 
$\dot{J_t}=\overline{\partial^2 S(z_t)\,J_t}$ 
together with Eq.~\eqref{flow}, 
whose cost is also $O(N^3)$. 
Figure \ref{fig:comp_cost} shows the real computation time 
for generating a single configuration, 
performed on a supercomputer (Yukawa-21) 
at Yukawa Institute, Kyoto University. 
We clearly see that it scales as expected, 
and smaller than  $O(N^{3-4})$ expected for the original TLTM. 
We also see that the aid of GPU is quite effective. 

The computational cost can be further reduced 
if we adopt an iterative method (such as BiCGStab)
as in Ref.~\cite{Alexandru:2017lqr}. 
The numerical cost is then expected to be $O(N^2)$ 
if the Krylov subspace iteration converges quickly. 
This factor will be multiplied by $O(N^{1/4})$ 
if we reduce the step size of molecular dynamics 
so that the acceptance rate of the final Metropolis test 
is independent of $N$.

\section{Summary and outlook}
\label{sec:summary_outlook}

We have reported that 
the tempered Lefschetz thimble method and its worldvolume extension, 
(WV-)TLTM, 
has a potential to be a reliable and versatile solution to the sign problem, 
because the algorithm solves the sign and ergodicity problems simultaneously 
and can be applied to any system in principle 
if it is formulated in a path-integral form with continuous variables.   
The (WV-)TLTM has been successfully applied to various models 
(yet only to toy models with small DOF at this stage), 
which include important toy models 
such as the Stephanov model (for finite density QCD), 
the 1D/2D Hubbard model (for strongly correlated electron systems), 
and the antiferromagnetic Ising model on the triangular lattice 
(for frustrated classical/quantum spin systems). 

We are now porting the code of WV-TLTM 
such that it can run on a large-scale supercomputer, 
which we expect to be completed soon. 
In parallel with this, 
it should be important to keep improving the algorithm itself 
so that the estimation can be made more efficiently 
for large-scale systems. 
It would be also interesting to combine various algorithms 
that have been proposed as solutions to the sign problem. 
An interesting candidate we have in mind as a partner of \mbox{(WV-)}TLTM 
is the tensor renormalization group method, 
which is actually complementary to Monte Carlo method 
in many aspects. 
A particularly important subject in the near future 
will be to establish a Monte Carlo algorithm 
for the calculation of time-dependent systems. 
This will open a way to the quantitative understanding 
of nonequilibrium processes, 
such as those happening in heavy ion collision experiments 
and in the very early universe. 

A study along these lines is in progress, 
and we hope we can report some of the achievements 
in the next Corfu conference.

\acknowledgments
The authors thank Issaku Kanamori, Yoshio Kikukawa and Jun Nishimura 
for useful discussions. 
M.F.\ thanks the organizers of Corfu 2021, 
especially George Zoupanos and Konstantinos Anagnostopoulos, 
for organizing wonderful conference series. 
This work was partially supported by JSPS KAKENHI 
Grant Numbers JP20H01900, JP21K03553. 
N.M.\ is supported by the Special Postdoctoral Researchers Program 
of RIKEN.
Some of our numerical calculations are performed 
on Yukawa-21 at Yukawa Institute for Theoretical Physics, 
Kyoto University.


\end{document}